\documentclass{PoS}

\usepackage{multirow}
\usepackage{amsmath}
\usepackage{amssymb}
\usepackage{slashed}
\usepackage{braket}
\usepackage{dsfont}
\usepackage{xifthen}

\title{\begin{picture}(0,0)(0,0)%
     \put(360,68){\makebox(0,0)[l]{\textnormal{\normalsize OU-HET-1032}}}%
\end{picture}Symmetries of the light hadron spectrum in high temperature QCD}

\ShortTitle{Symmetries of the light hadron spectrum}
\author{\speaker{C. Rohrhofer}$^a$\thanks{E-mail: \texttt{crohrhofer@het.phys.sci.osaka-u.ac.jp}},
        Y.~Aoki$^b$,
        G.~Cossu$^c$,
        H.~Fukaya$^a$,
        C.~Gattringer$^d$,
        L.Ya.~Glozman$^d$,
        S.~Hashimoto$^e$$^f$,
        C.B.~Lang$^d$,
        K.~Suzuki$^g$
        \\
        \\
        $^a$ Department of Physics, Osaka University, Toyonaka 560-0043, Japan\\
        $^b$ RIKEN Center for Computational Science, Kobe 650-0047, Japan\\
        $^c$ School of Physics and Astronomy, The University of Edinburgh, Edinburgh EH9 3JZ, United Kingdom\\
        $^d$ Institute of Physics, University of Graz, 8010 Graz, Austria\\
        $^e$ KEK Theory Center, High Energy Accelerator Research Organization (KEK), Tsukuba 305-0801, Japan\\
        $^f$ School of High Energy Accelerator Science, The Graduate University for Advanced Studies (Sokendai), Tsukuba 305-0801, Japan\\
        $^g$ Advanced Science Research Center, Japan Atomic Energy Agency (JAEA), Tokai 319-1195, Japan\\   
        }  
                
\abstract{Properties of QCD matter change significantly around the chiral\
          crossover temperature, and the effects on $U(1)_A$ and topological\
          susceptibilities, as well as the meson spectrum have been studied\
          with much care. Baryons and the effect of parity doubling in this\
          temperature range have been analyzed previously by various other\
          groups employing different setups. Here we construct suitable\
          operators to investigate chiral and axial $U(1)_A$ symmetries in\
          the baryon spectrum. Measurements for different volumes and\
          quark-masses are done with two flavors of chirally symmetric\
          domain-wall fermions at temperatures above the critical one.\
          The possibility of emergent $SU(4)$ and $SU(2)_{CS}$ symmetries is\
          discussed.}

\FullConference{37th International Symposium on Lattice Field Theory - Lattice2019\\
		16-22 June 2019\\
		Wuhan, China}

\begin{document}

\section{Introduction}

While in the vacuum $SU(2)_L \times SU(2)_R$ chiral symmetry of QCD is
spontaneously broken by a finite expectation value of the quark
condensate $\langle \bar q q \rangle$, and thus realized in the
Nambu--Goldstone mode, this behaviour changes with increasing temperatures.
For physical quark masses the condensate rapidly drops at a pseudo-critical
temperature $T_c$~\cite{Borsanyi:2010bp,Bazavov:2011nk}
and marks a general QCD transition temperature above which
$SU(2)_L \times SU(2)_R$ chiral symmetry is restored.
Anomalous breaking of the axial $U(1)_A$ symmetry and associated complications in
measurements make a precise lattice analysis of its fate more intricate.
However, there is good evidence that its breaking effects are strongly
suppressed after the chiral transition:
by studying the eigenvalue spectrum of the Dirac operator, flavor
non-singlet meson susceptibility differences and topological susceptibility,
the JLQCD collaboration has found effective $U(1)_A$ restoration in the
chiral limit around $T \sim 200$~MeV in a carefully constructed chirally
symmetric setup of lattice simulation~\cite{Tomiya:2016jwr,Suzuki:2019vzy}.

In this work we study the chiral symmetry properties of the light hadron
spectrum in two-flavor QCD at a temperature of $T \sim 220$ MeV.
For mesons similar calculations~\cite{Brandt:2016daq,Bazavov:2019www}
have found a symmetry pattern in agreement with a restoration of both
symmetries above 200 MeV.
In the baryon sector various studies have found signals of parity doubling
for nucleons at high temperature~\cite{DeTar:1987xb,Datta:2012fz,Aarts:2015mma}
by comparing different parity states of the same interpolating field.
Here we use meson and nucleon interpolators with different chiral
transformation properties to explicitly check for
$U(1)_A$ and $SU(2)_L \times SU(2)_R$ restoration,
as well as parity doubling for nucleons.

\section{Method}

Central elements for studying the chiral properties of hadrons are local
interpolating fields with different chiral transformation properties.
To summarize the meson measurement setup detailed
in~\cite{Rohrhofer:2017grg,Rohrhofer:2019qwq},
we list the pairs of flavor nonsinglet operators connected by
$U(1)_A$ and $SU(2)_L \times SU(2)_R$ (in the following abbreviated as $SU(2)_A$)
transformations:
\begin{align}
  \;\; U(1)_A&:\qquad PS \;\; (u \; \gamma_5 \; d) \longleftrightarrow \;\; S \;\;\; (u \; \mathds{1}       \; d),
  \label{equ:mesonU1}\\
  \;\;SU(2)_A&:\qquad V_k \;\;\; (u \; \gamma_k \; d) \longleftrightarrow \;\; A_k \; (u \; \gamma_k\gamma_5 \; d),
  \label{equ:mesonSU2A}
\end{align}
where in this study we use the vector components $k=1,2$.  
In the baryon sector we probe nucleons
\begin{equation}
N_i = \epsilon_{abc} \left( u^T_a \Gamma_1 d_b \right) \Gamma_2 u_c,
\label{equ:interpolatingfield}
\end{equation}
where the gamma structures $\Gamma_1$ and $\Gamma_2$ are given in Table \ref{tab:ops}.
\begin{table}[b]
\centering
\begin{tabular}{ccccl}
\hline\hline
\rule{0pt}{3ex}
  Abbreviation             &
  $\Gamma_1$               &
  $\Gamma_2$               &
  Chiral-Parity Group Rep. &
  Symmetries
\\[1ex]\hline\rule{0pt}{3ex}
 %%%%%%%%%%%
$N_1$ & $C\gamma_5$         & $\mathds{1}$   &   $[(0,\frac{1}{2})+(\frac{1}{2},0)]_a$ &   \multirow{2}{1cm}{$\left.\begin{aligned}\\ \end{aligned}\right] U(1)_A$} \\
$N_2$ & $C$                 & $\gamma_5$     &   $[(0,\frac{1}{2})+(\frac{1}{2},0)]_b$ &   \\[1ex]
\hline \rule{0pt}{3ex}
$N_3$ & $C\gamma_4\gamma_5$ & $\mathds{1}$   &   $(\frac{1}{2},1)+(1,\frac{1}{2})$     &   \multirow{2}{1cm}{$\left.\begin{aligned}\\ \end{aligned}\right] SU(2)_A$}\\
$N_4$ & $C\gamma_4$         & $\gamma_5$     &   $(\frac{1}{2},1)+(1,\frac{1}{2})$     &   \\[1ex]
\hline\hline
\end{tabular}
\caption{Nucleon operators used in this study. The second and third column
         give the gamma structures used in (\ref{equ:interpolatingfield}).
         The last column shows which symmetry transformation connects the
         different operators~\cite{Cohen:1996sb,Denissenya:2015woa}.}
\label{tab:ops}
\end{table}
The operators (\ref{equ:interpolatingfield}) mix both parity states and require proper
projection to positive or negative parity:
\begin{equation}
N^\pm_i = \frac{(1\pm \gamma_4)}{2}N_i.
\label{equ:parityprojection}
\end{equation}
For measurements of the screening spectrum in $z$-direction we use the
projection~(\ref{equ:parityprojection}) with $\gamma_3$ instead of $\gamma_4$~\cite{Datta:2012fz}.
Furthermore, the momentum projection is done onto the lowest Matsubara frequency $\omega_0=\pi T$,
which is non-zero due to the antiperiodic boundary conditions in $t$-direction~\cite{DeTar:1987xb}:
\begin{equation}
C(n_z)=
\sum\limits_{n_x,n_y,n_t}
e^{in_t \omega_0}
\langle
N^\pm(n_x,n_y,n_z,n_t)
\bar N^\pm(\mathbf{0})
\rangle.
\end{equation}

The gauge ensembles used in this study are part of an ongoing large scale
simulation by the JLQCD collaboration. 
Its setup is described in detail in \cite{Cossu:2015kfa,Tomiya:2016jwr}.
Here we use a single temperature of $T=220$ MeV with $N_t=12$ as the lattice
extent in time direction, and an inverse gauge coupling $\beta=4.30$
corresponding to a lattice cutoff $1/a=2.6$ GeV.
Two flavors of dynamical domain-wall fermions are used with an auxiliary
5-th dimension of $L_s=16$, which ensures a violation of the Ginsparg-Wilson
condition by less than $1$~MeV.
The bare quark masses $m_u = m_d \equiv m_{ud}$ vary between $0.01-0.001$
corresponding to $26-2.6$ MeV, where the lightest mass is slightly below the physical value.
Calculations are performed on four different volumes with spatial extent between
$N_s=24-48$. This results in an aspect ratio $N_s/N_t=2$ for our smallest volume
and $N_s/N_t=4$ for our largest volume.
Measurements on each ensemble include at least $\mathcal{O}(50)$ independent
configurations.

To compare results of differently sized lattices when presenting results for the
screening spectrum we give spatial distances as the
dimensionless combination
\begin{align}
z\,T \; = \; (n_z a)/(N_t a) \; = \; n_z/N_t \; ,
\label{z_dimless}
\end{align}
where~$z$ is the physical distance and $n_z$ the distance in lattice units.

\section{Results}

In a related previous study we found restoration of both, $U(1)_A$ and
$SU(2)_L \times SU(2)_R$, chiral symmetries in the spectrum of
mesons above a temperature of $T=220$~MeV~\cite{Rohrhofer:2017grg,Rohrhofer:2019qwq}.
We measured this by calculating ratios of correlation functions for pairs of operators,
which are connected by symmetry transformations.
A ratio converging to 1 with decreasing quark mass signals identical behaviour
of the different correlators and derived quantities, e.g. screening masses,
and therefore the restoration of symmetry.
Here we take a more direct approach and
compare the differences of local screening masses, the logarithmic slopes at a
certain distance,
evaluated at a distance $zT$ for a single temperature $T=220$ MeV.

\begin{figure}
  \centering
  \includegraphics[scale=0.58]{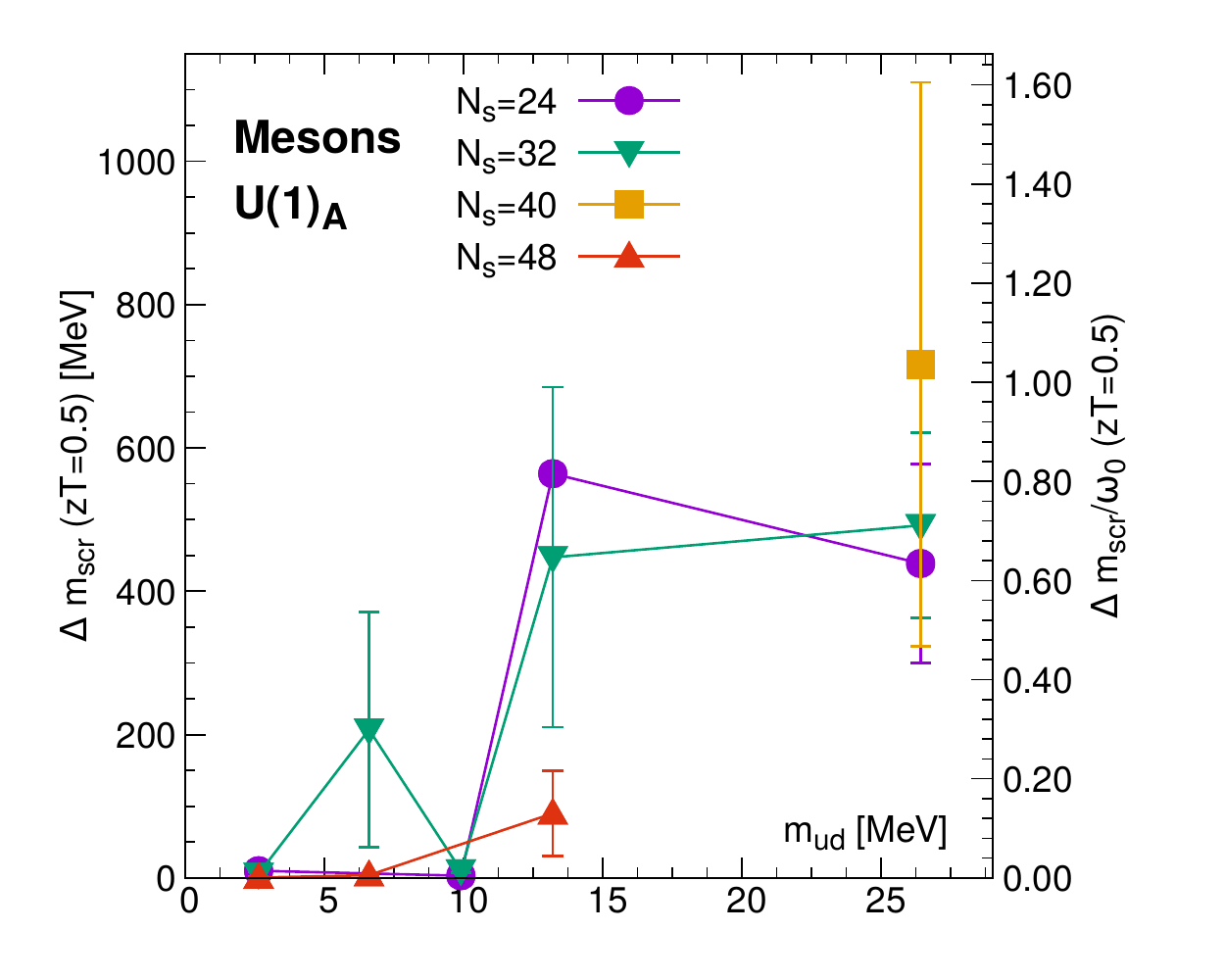}
  \hspace{0cm}
  \includegraphics[scale=0.58]{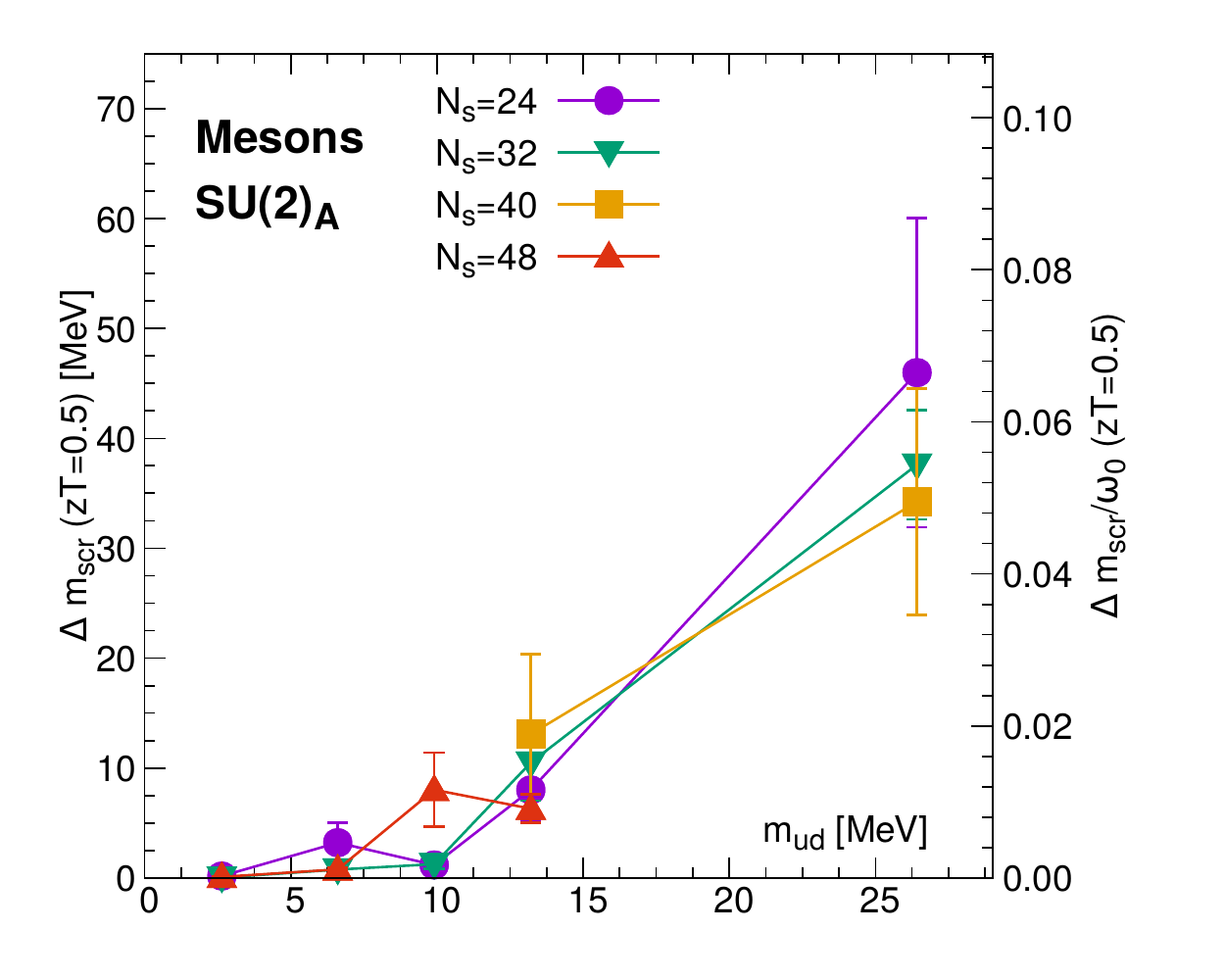}
  \caption{\textit{Left:} $U(1)_A$ symmetry breaking in the meson screening spectrum as
           measured by (\ref{equ:MDU1}).
           \textit{Right:}~chiral symmetry breaking  of (\ref{equ:MDSU2A}).
					 The mass difference is evaluated at $zT=0.5$ for all lattices.}
  \label{fig:mesons}
\end{figure}
Hence, for mesons the breaking of $U(1)_A$ is measured through
\begin{equation}
  \;\;U(1)_A:\qquad \Delta m_{scr} = |m^{PS}_{scr} - m^{S}_{scr}|,
\label{equ:mesonMDU1}
\end{equation}
which is shown on the left of Figure~\ref{fig:mesons}.
Flavor non-singlet $SU(2)_A$ symmetry breaking 
\begin{equation}
  \;\;SU(2)_A:\qquad \Delta m_{scr} = |m^{V_k}_{scr} - m^{A_k}_{scr}|
\label{equ:mesonMDU1}
\end{equation}
is shown on the right side of Figure~\ref{fig:mesons}.
Both show a strong quark mass dependence, while 
the $U(1)_A$ data is significantly noisier.
Different lattice volumes
do not seem to affect the symmetry breaking substantially.
The difference in masses for both cases is less than 5 MeV at the lightest quark mass,
which is less than 2 \% of the temperature.
This agrees with the conclusions from our previous measurements in
\cite{Rohrhofer:2017grg,Rohrhofer:2019qwq}.

The situation for baryons is summarized in Table~\ref{tab:ops}:
Operator $N_1$ belongs to a $[(0,1/2)\oplus (1/2,0)]$ representation of the
chiral-parity group.
Operator $N_2$ belongs to a different $[(0,1/2)\oplus (1/2,0)]$ representation
of the same chiral-parity group.
Both representations have the same isospin content and are related by axial
$U(1)$ rotations.
Breaking of $U(1)_A$ is thus measured through
\begin{equation}
\;\;U(1)_A:\qquad \Delta m_{scr} = |m^{N_1^+}_{scr} - m^{N_2^+}_{scr}|.
\label{equ:MDU1}
\end{equation}
$SU(2)_A$
on the other hand mixes operators within a multiplet of the chiral-parity
group, here $N_3$ and $N_4$ of $(1/2,1) \oplus (1,1/2)$.
Its breaking is measured through
\begin{equation}
SU(2)_A:\qquad \Delta m_{scr} = |m^{N_3^+}_{scr} - m^{N_4^+}_{scr}|.
\label{equ:MDSU2A}
\end{equation}
\begin{figure}
  \centering
  \includegraphics[scale=0.58]{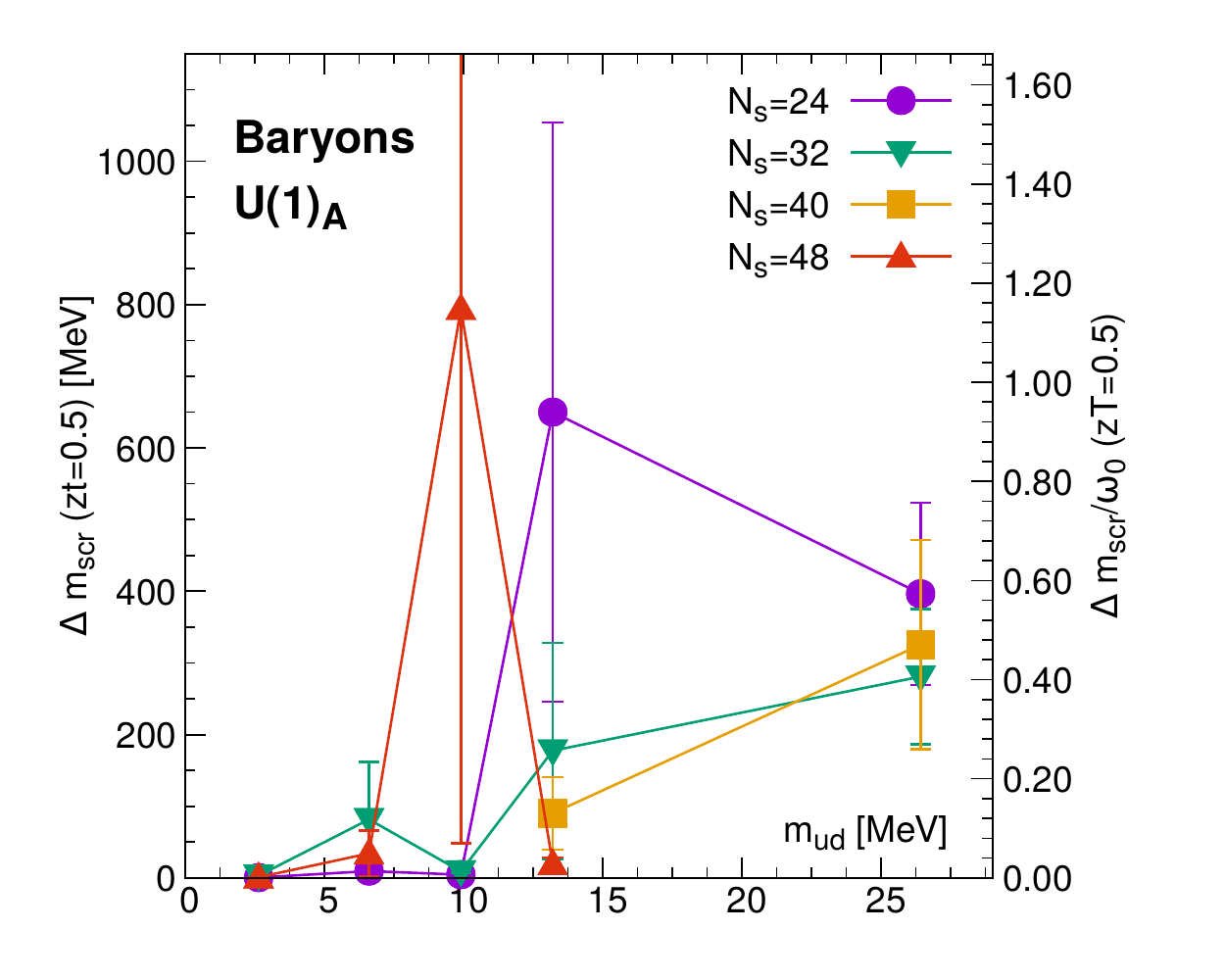}
  \hspace{0cm}
  \includegraphics[scale=0.58]{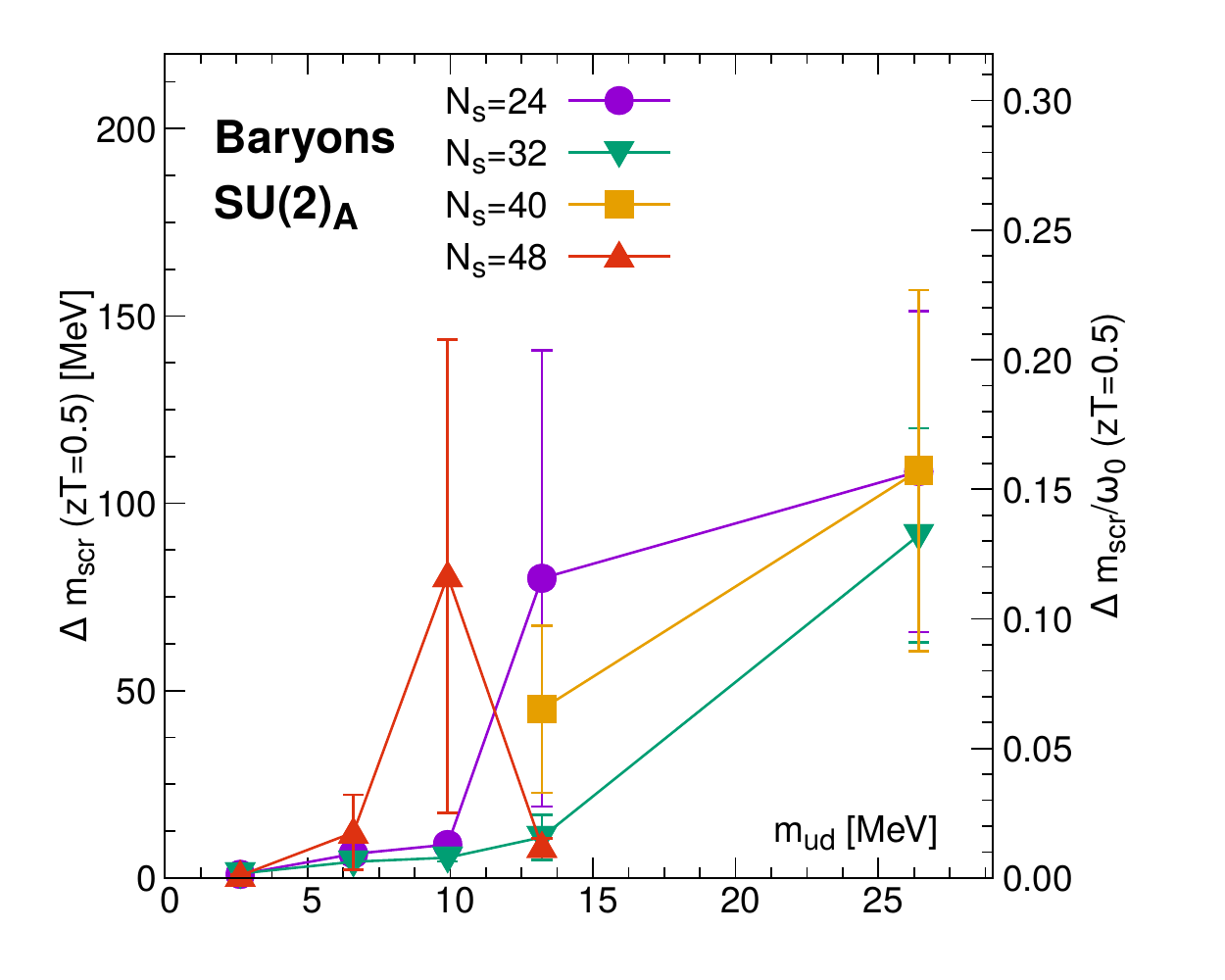}
  \caption{\textit{Left:} $U(1)_A$ symmetry breaking in the baryon screening spectrum as
           measured by (\ref{equ:MDU1}).
           \textit{Right:}~chiral symmetry breaking  of (\ref{equ:MDSU2A}).
					 The mass difference is evaluated at $zT=0.5$ for all lattices.}
  \label{fig:baryons}
\end{figure}
Figure~\ref{fig:baryons} shows the results for baryons at a temperature of $T=220$ MeV.
The left panel shows $U(1)_A$ breaking with a quark mass dependence and noise similar to the mesonic case in Fig.~\ref{fig:mesons}.
The effective mass difference at our lightest quark mass is less than 5 MeV for all volumes,
which is $2\%$ of the temperature.
The right panel shows $SU(2)_A$ breaking: although the violation at heavier quark masses is slightly higher than for mesons,
it shows the same qualitative behaviour towards the chiral limit again
with a mass difference of less than 5 MeV for the lightest quark masses.
We note that especially for heavier quark masses the signal shows strong fluctuations.

As each multiplet of the chiral-parity group includes the parity partners
of operators by construction,
the restoration of $SU(2)_A$ can be seen as signal for parity restoration
and vice versa.
These results thus
imply parity doubling at a temperature of~$220$ MeV.
The studies mentioned in the introduction, however, measure parity doubling
by comparing the parity partners of the same interpolating field, namely $N_1$.
In~\cite{Datta:2012fz,Aarts:2015mma} a specially constructed $R$-value
\begin{equation}
R_{integrated}=\frac
{\sum_{t=1}^{{N_t}/2} R(t)/\sigma^2(t)}
{\sum_{t=1}^{{N_t}/2} 1/\sigma^2(t)},
\label{equ:Rint}
\end{equation}
is used to quantify the level of parity breaking for one lattice ensemble.
It is a summation of
\begin{equation}
R(t)=\frac
{N_1^+(t)-N_1^+(N_t-t)}
{N_1^+(t)+N_1^+(N_t-t)},
\label{equ:Rvalue}
\end{equation}
along half of the lattice
(the other half is the same by construction), weighted by the statistical
uncertainty $\sigma$.
Eq.~(\ref{equ:Rvalue}) describes the asymmetry of the correlation function for a
specific parity projected operator ($N_1^+$ in this case) at time slice $t$.
For both expressions (\ref{equ:Rvalue}) and (\ref{equ:Rint}) we
exchange $t\rightarrow n_z$ and $N_t\rightarrow N_z$ in case of
measurements in $z$-direction.

\begin{figure}[b]
  \centering
  \includegraphics[scale=0.55]{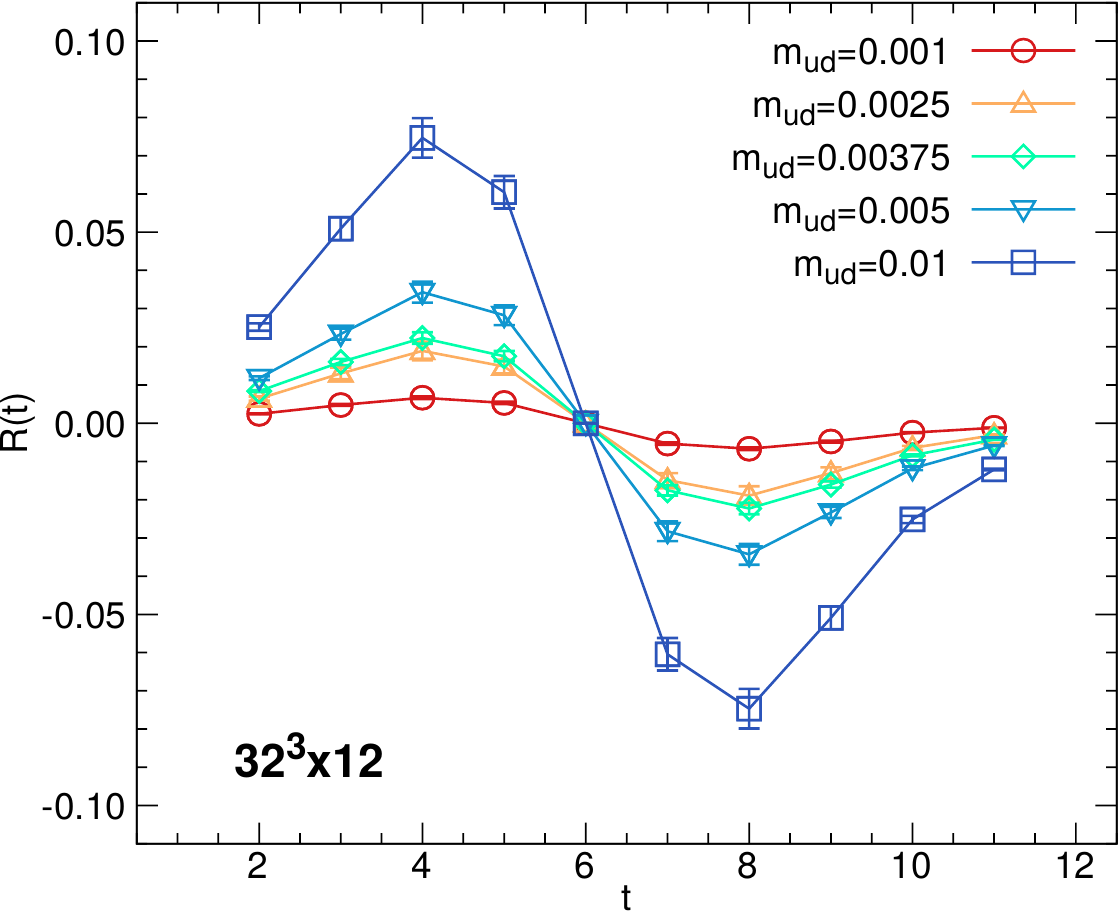}
  \hspace{1cm}
  \includegraphics[scale=0.55]{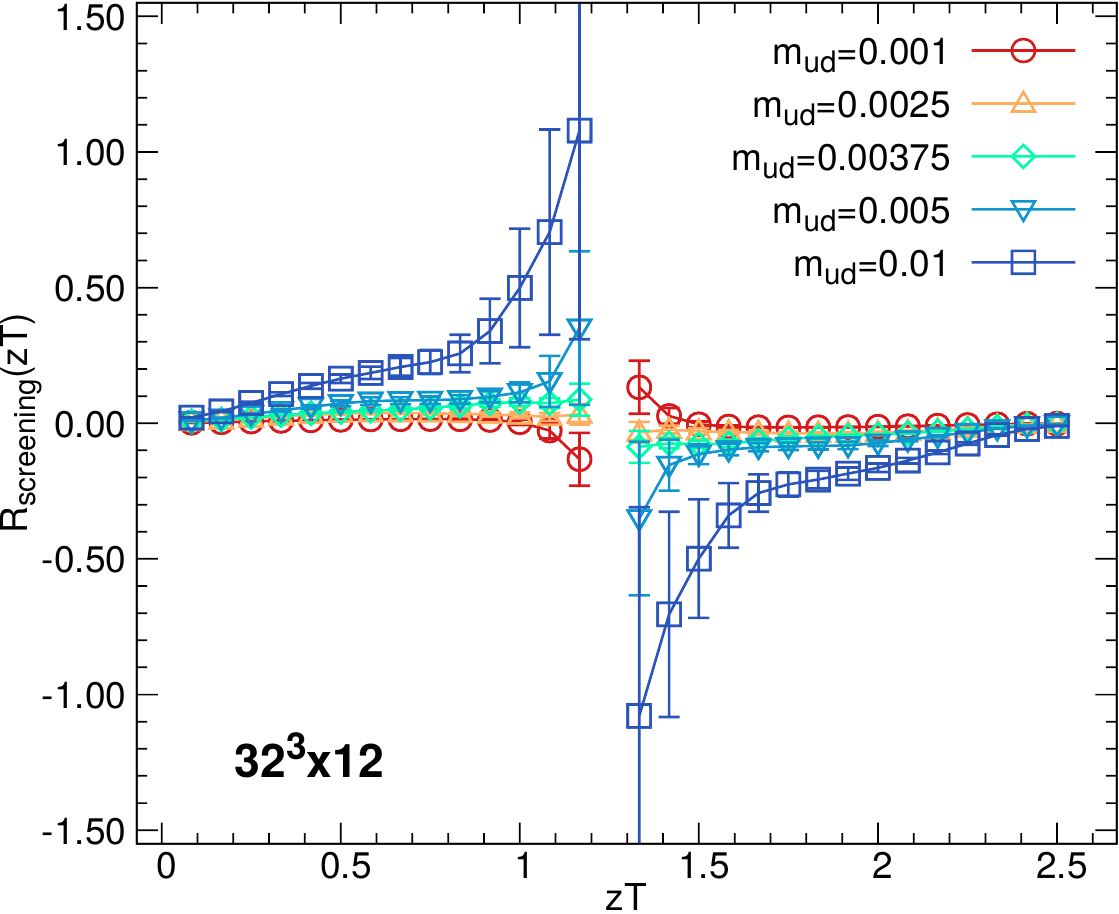}
  \caption{R-value (\ref{equ:Rvalue}) for the nucleon spectrum measured in
           $t$-direction (left) and the corresponding screening spectrum
           measured in $z$-direction (right).}
  \label{fig:R}
\end{figure}
On the left side of Figure~\ref{fig:R} the $R$-value~(\ref{equ:Rvalue})
is shown for five different quark masses on a $32^3\times 12$ lattice,
measured in $t$-direction. While the numerical value is meaningless
(due to the missing renormalization of our lattice operators), the
quark mass dependence of $R$ is significant.
Decreasing quark mass strongly suppresses the $R$-value, and hence the
asymmetry in correlators of positive and negative parity states.

The right side of Figure~\ref{fig:R} shows the corresponding $R$-value in
the screening spectrum of the same ensembles. A sign change of
screening correlators around the midpoint of the lattice causes the data
to behave differently in this region, the overall interpretation stays the
same. Also in this case a strong quark mass dependence can be observed.

\begin{figure}[t]
  \centering
  \includegraphics[scale=0.67]{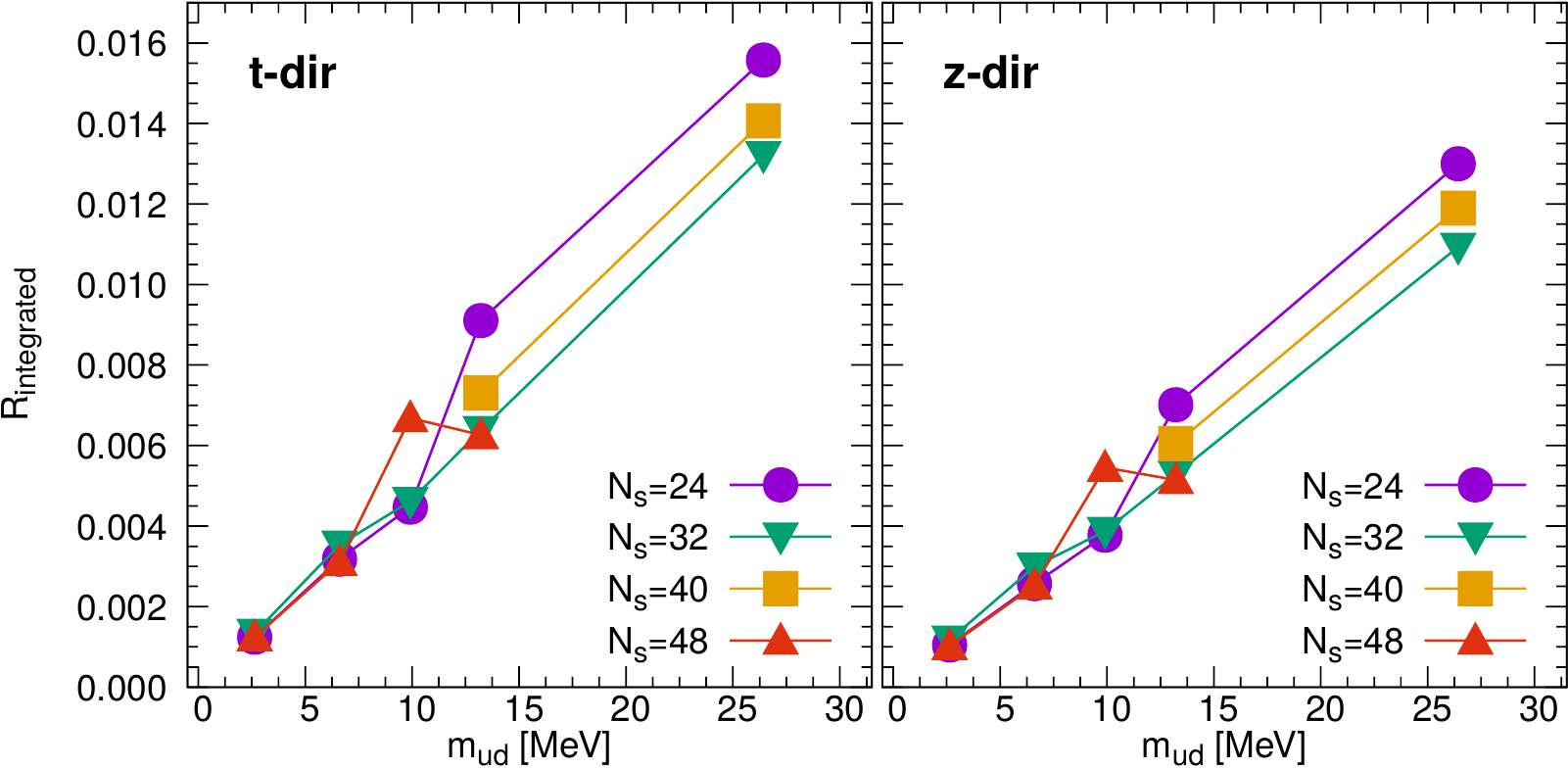}
  \caption{Integrated R-value (\ref{equ:Rint}) for different quark masses and volumes.
 					 The left panel shows the nucleon spectrum measured in $t$-direction, the
					 right panel its screening spectrum measured in $z$-direction.}
  \label{fig:Rint}
\end{figure}
Figure~\ref{fig:Rint} finally shows the integrated $R$-values~(\ref{equ:Rint})
for all ensembles in this study --- on the left side for $t$-direction measurements,
on the right side for the screening spectrum.
Although the approach~(\ref{equ:Rint}) lacks an estimate of uncertainty, the
data so far is suggestive that parity restoration is happening in the chiral limit,
in agreement with the findings earlier in this section.

\section{Summary}

In this work, using two flavors of chirally symmetric quarks,
we studied the restoration of chiral and $U(1)_A$ symmetry
in the spectrum of light mesons and nucleons above the chiral transition at
a temperature $T=220$ MeV.
We calculated the difference of local screening masses
for symmetry partners at $zT=0.5$,
and found that in all cases the mass difference decreases significantly with lighter
quark masses.
At our lightest quark mass of $2.6$ MeV the breaking is less than
5 MeV, or 2\% of the temperature.
These results are in agreement with our previous findings for the meson
spectrum in~\cite{Rohrhofer:2017grg,Rohrhofer:2019qwq},
and suggest restoration of both symmetries in the chiral limit at~$T=220$~MeV.

For nucleons the effect of parity doubling was investigated at the same temperature
in the chiral limit, for both $t$-correlators and the screening spectrum,
and the results suggest the emergence of parity doubling.

Our work additionally shows the emergence of a $SU(2)_{CS}$ \textit{chiral spin}
and $SU(4)$ symmetry in the meson spectrum of QCD at high temperature,
which becomes manifest upon inclusion of the six tensor elements of flavor non-singlet
meson operators $u \sigma_{\mu\nu} d$.
Albeit not presented here, detailed results and physical consequences are discussed
in~\cite{Rohrhofer:2019qwq}.

\newpage

\section*{Acknowledgments}
Support from the Austrian Science Fund (FWF) through the grants
DK W1203-N16 and P26627-N27, as well as from NAWI Graz is acknowledged.
This work was supported in part by JSPS KAKENHI Grant Numbers JP18H03710,
JP18H01216, JP18H04, and by MEXT as ``Priority Issue on post-K computer''.
Numerical simulations are performed on Oakforest-PACS at JCAHPC under
a support of the HPCI System Research Projects (Project IDs: hp170061,
hp180061 and hp190090),
the Multidisciplinary Cooperative Research Program in CCS, University of
Tsukuba (Project IDs:xg17i032, xg18i023),
and on the IBM System Blue Gene Solution at KEK
under a support of its Large Scale Simulation Program (No. 16/17-14).
This work is supported in part by the Post-K supercomputer project through
the Joint Institute for Computational Fundamental Science (JICFuS).

\end{document}